# Progress towards 3D-printing diamond for medical implants: A review

Aaqil Rifai,[a,#] Shadi Houshyar,[a] and Kate Fox[a,b,#]

a) School of Engineering, RMIT University, VIC Australia

b) Centre for Additive Manufacturing, RMIT University, VIC, Australia

[#] Corresponding author email: aaqil.rifai@rmit.edu.au and kate.fox@rmit.edu.au

Abbreviations

[1]Additive Manufacturing

[2]Selective Laser Melting

[3]Chemical Vapour Deposition

[4]Nanodiamonds

[5]High Pressure High Temperature

---

[1] AM
[2] SLM
[3] CVD
[4] NDs
[5] HPHT

# Abstract

Additive manufacturing or 3D-printing is used to create bespoke items in many fields, such as defence, aerospace and medicine. Despite the progress made in 3D-printed orthopaedic implants, significant challenges remain in terms of creating a material capable of osseointegration while inhibiting bacterial colonisation of the implant. Diamond is rapidly emerging as a material with an extensive range of biomedical applications, especially due to its excellent biocompatibility. However, diamond is a difficult material to fabricate, owing to its extreme level of hardness and its brittleness. New methods of fabrication including additive manufacturing, have overcome some of these challenges and given rise to an increase in the use of diamond-based implants in both soft and hard tissue applications. Therefore, due to the unique properties of diamond, it is being considered as a facilitator of bone growth and subsequent tissue integration. This review outlines the recent progress in fabricating diamond for orthopaedic application, specifically focusing on the different fabrication approaches and their applicability *in vitro* and *in vivo*. The prospects and challenges for using diamond in medical implant technologies are also discussed.

# Introduction

Recently, the manufacture of implantable materials has transitioned from casting and machining [1, 2] to a customisable approach known as additive manufacturing (AM) or three-dimensional (3D) printing [3, 4]. A popular technique to manufacture metals is to use selective laser melting (SLM), which has revolutionised the manufacturing of many devices, particularly implants. For implant production, these techniques offer efficient use of raw materials with minimal waste, satisfactory geometric accuracy, and robust mechanical integrity. Typically, the metals that are used in medical applications include titanium, stainless steel, carbon nitride, zirconium oxide, titanium oxide, titanium nitride and cobalt-chromium alloys [5, 6]. Compared to traditional methods of manufacturing implants, AM allows complex structures to be printed on a layer-by-layer basis. Additive manufacturing offers a great advantage in the development of personalised healthcare products that are customised to the specific needs of a patient; particularly in regard to the size, shape and internal structural features of the implant.

The average lifetime of an implantable prosthesis, such as a hip implant is between 5 to 25 years. These implants and devices are widely used as permanent or temporary substitutes for joints or bones [7, 8]. The primary goal of orthopaedic implants is to provide a mechanical fixture that can withstand the loads carried by the original host bone. Implants facilitate pain relief and promote rehabilitation compared with having no implant. However, orthopaedic implants are limited in their durability before they fail due to the lack of osseointegration and aseptic loosening [9-11]. Wear debris can cause aseptic loosening when sliding forces are created between the implant and bone. Commonly, smaller wear particles are engulfed by macrophages and multinucleated giant cells, and the larger debris is comprised of fibrous tissue. This response can influence the release of cytokines and cause inflammation. It results in osteolysis (bone degradation) and failure, due to the loosening of the implant [5]. In other cases, orthopaedic fixative devices may aide the mechanical and biological functions to some degree but succumb to bacterial infections. This can lead to implant failure and result in revision surgery [12-14]. Hence, alternative materials are continuously being investigated to mimic the properties of bone and provide a better interface of the cell with material interaction.

The application of a secondary coating on an implant can improve the biological interaction between the implant and the surrounding tissue [15]. However, orthopaedic implant coatings can be challenging for many reasons: the complexity of the underlying structures, and the coating parameters that affect the adhesion, thickness, roughness and morphology. As a result, these parameters require optimisation. Ceramic, carbon and metallic coatings are some of the commonly used coatings for metal implants [16-18]. Ceramic coatings, including hydroxyapatite (HA) can improve the bone to implant interface by promoting bone-in and outgrowth [17]. However, the long-term stability and biodegradability of HA coatings are of concern [19]. In other instances, silver is used as a coating to provide antibacterial effects on an implant [20]. Although silver is known for its antibacterial properties, it is not typically used to enhance mammalian cell growth simultaneously. Furthermore, various types of coatings can be used to improve the hydrophilicity, hydrophobicity and corrosion resistance of implants. These include polymeric, minerals and nanocomposite coatings. Likewise, several coating methods exist, such as sol-gel deposition, sputter coating, spin coating, dip coating, chemical vapour deposition, thermal spray coating, electrostatic spray coating and pulsed laser deposition [15, 21, 22]. An ineffective coating typically leads to the failure of the implant. The lack of a uniform and well-adhered surface coating to the implant encumbers osseointegration and increases the likelihood of bacterial infection.

Diamond provides several solutions for biomedical implants due to its unique properties including hardness, mechanical stability and biocompatibility [23]. In the recent decades, carbon-based materials have gained interest as they demonstrate significant benefits for orthopaedic and dental implantation [24, 25]. The fascination with using diamond in implants arose in applications of wear-resistant hip joints [24] and, with time, diamond has become a key material used in orthopaedics, dental and cardiovascular engineering,[24, 26] primarily due to its biocompatibility with the human body. Chemical vapour deposition (CVD) diamond has been used as a coating for mandibular plates,[27, 28] heart valves,[29] neural stimulators,[30] and joint replacements [31]. Likewise, diamond particles also exhibit excellent properties in medicine [32]. Wear-resistance and biocompatibility are two of the most desired characteristics for implant materials.

This review will first introduce conventional diamond, then focus more on the characteristics and fabrication of diamond-based scaffolds, assessment of the bio-interface both *in vitro* and *in vivo* to understand the biocompatibility and osseointegration of diamond-based scaffold, and finally details on how this material can be translated into 3D printing.

**Fabrication of diamond: synthesis and types**

The fascination of diamond can be traced back thousands of years to when its unique physical and optical properties were first observed in gemstones. Around the 1950s, the first synthetic diamonds were produced, leading to commercialisation in the same decade. These types of diamond were known as high-pressure high-temperature (HPHT) diamonds [33]. They were initially used to provide high-wear-resistance grit for cutting tools. Later, the fabrication of diamond transitioned to low-pressure growth using chemical vapour deposition. Unlike diamond produced with HPHT, chemical vapour deposition (CVD) diamond can be grown conformably on some surfaces as a microcrystalline or nanocrystalline film or in a polycrystalline form. Diamond is mainly applied as a coating using CVD techniques involving plasma containing hydrogen and carbon for deposition [34]. To allow CVD to nucleate into a thin film, a feed-gas containing carbon, and diamond seeds on the surface of coated material, are required. The presence of hydrogen atoms in the gas mixture is what mediates the creation of diamond over other carbon forms such as graphite. The graphitic material is etched away, allowing the diamond to grow. Methane is typically used to promote the growth of diamond; however, several sources exist [35, 36]. The two most widely used CVD fabrication methods are hot-filament and plasma-enhanced CVD (Fig. 1). In hot-filament CVD, there is a thin, high-melting-temperature wire positioned close to the diamond seeds. A feed gas mixture is presented, and the wire is heated to 2000–2500 °C whilst the gas mixture is activated, and diamond seeds are heated. In comparison, the plasma-enhanced CVD method uses a microwave energy source fixated into the gas mixture to develop a high-energy plasma close to the diamond seeds [34, 37]. However, problems exist with certain materials and diamond due to the thermal mismatch, causing delamination.

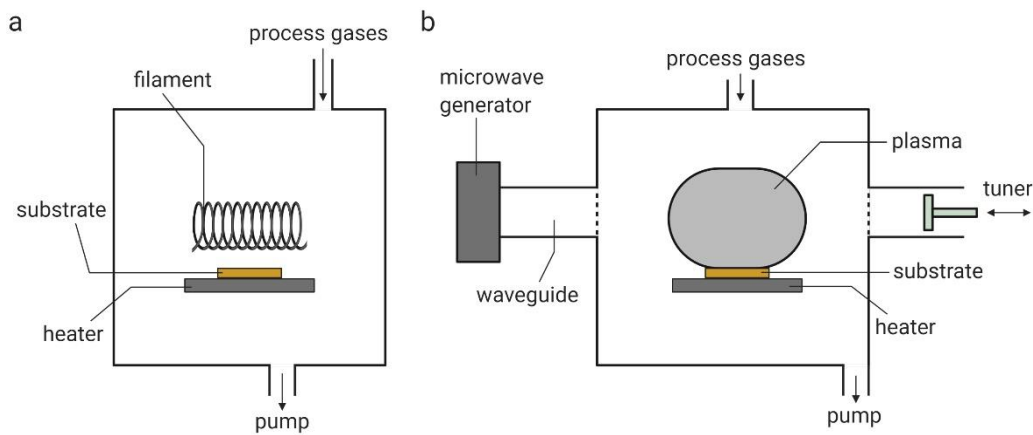

**Figure 1.** Schematic representation of two types CVD reactors. (a) hot filament and (b) plasma-enhanced microwave plasma reactor.

On the other hand, nanodiamonds (NDs) can be used as seeds and are generated either by a top-down approach (milling a larger diamond) or bottom-up approach (TNT detonation in a pressurised container). Other than their use in diamond coatings, NDs themselves are excellent materials in medicine. For example, fluorescent NDs have successfully demonstrated their capabilities as biomarkers for applications in cellular tracking [38]. The particle form with this method is known to be less expensive than the CVD method, providing new avenues and more freedom of use. Table 2 shows a list of the main forms of diamonds and their commonly used applications.

**Table 2**. Types of diamond used in biomedicine [37].

| Type | Crystal Size | Appearance | Biomedical Use |
| --- | --- | --- | --- |
| Single crystals | µm – cm | Clear - coloured | - No current biomedical application but possibilities in sensing |
| Microcrystalline diamond films | 0.5 - 50 µm | Black - clear | - Implantable electrodes<br>- Implant encapsulation<br>- Heat dissipation<br>- Wear resistance |
| Nanocrystalline diamond | 10 - 500 nm | Black | - Wear resistance<br>- Anti-corrosion barrier films<br>- Barrier film (Biomedical) |
| Ultrananocrystalline diamond | 5 – 10 nm | Black | - Implantable electrodes<br>- Wear resistance<br>- Barrier film (Biomedical) |

| Nanodiamond | 5 - 100 nm | Black | - Antibacterial coatings |
| | | | - Sensing |
| | | | - Drug delivery |

More recently, diamond is showing potential to be 3D-printed with composite materials. Whilst 3D printed diamond scaffolds are limited, diamond and metal or diamond and polymer mixtures are becoming more prevalent, especially for applications in wear resistance. The 3D-printing technologies include stereolithography and laser metal deposition. This is discussed further in sections below.

**Diamond in medicine**

Chemical vapour deposition diamond

Examples of current medical implant applications using CVD diamond coatings include heart valves, neural stimulators, mandibular plates, and joint replacements where the interaction between the surrounding cells and device is necessary [20, 28, 39]. The use of diamond can be somewhat restrictive by its inherent properties, such as its extreme hardness in load-bearing applications (e.g. femoral stems). As a result, the hardness of diamond restricts the contact surfaces at which mechanical wear may occur or where an implant requires the material to guard against corrosion. However, diamond has found a role in medicine where biocompatibility and biointerfacing are important. The use of CVD diamond is more common with metals such as cobalt-chromium and stainless steel [40]. Typically, in total replacement hip surgeries, cobalt-chromium alloys are used, due to its superior corrosion resistance, mechanical qualities and biodegradation capacity. However, cobalt-chromium alloys are prone to the release of metal ions under prolonged wear [31, 41]. Diamond can limit the release of toxic ions when it is applied as a coating, potentially increasing the longevity of the implant. Although diamond offers a number of opportunities for biomedical use, progress in developing diamond and diamond-coated implants remains slow.

Nanodiamond

As with CVD diamond, NDs exhibit favourable characteristics in biomedical applications due to their enhanced cell binding property. They are also low cost with great fluorescent capability and biocompatibility [32, 42, 43]. Common biomedical uses of NDs include drug delivery,[44] biosensing,[45] and antibacterial [46] applications. They are highly adaptable and are able to deliver antigens, antibodies, nucleic acids and imaging agents into the specified cells, where the diagnostic or therapeutic molecules are released and utilised [47, 48]. The properties of NDs provide a bridge between quantum physics and biology, showing the capacity to track nitrogen vacancies within the diamond lattice *in vivo* [42, 46, 49, 50].

In a study by Lechleitner *et al.*,[51] borosilicate glass samples coated with nanoparticulate diamond powder provided improved scaffolds for epithelial cell adhesion using a modification in surface chemistry. The termination of ND powders and films can dictate the epithelial response. The differences between the diamond surfaces can be characterised in terms of the electrical conductivities, negative and positive electron affinity and the degree of hydrophobicity and hydrophilicity. The variations in the terminated surfaces are primarily due to the differences between the surface dipole moments of carbon/oxygen and carbon/hydrogen bonds [52]. The control of a material's wettability is important to mediate the cell response on the surface [20, 53, 54].

In another report, Chow *et al.* show that NDs are biocompatible *in vivo* as a potential drug delivery platform (Fig. 2). The surface of the NDs enables table drug sequestering, which resulted in a 10-fold increase in the circulation of blood. This complex of ND and doxorubicin surpassed drug efflux and substantially increased apoptosis and tumour growth inhibition compared to traditional doxorubicin treatment in mammary carcinoma and murine liver tumour models [55].

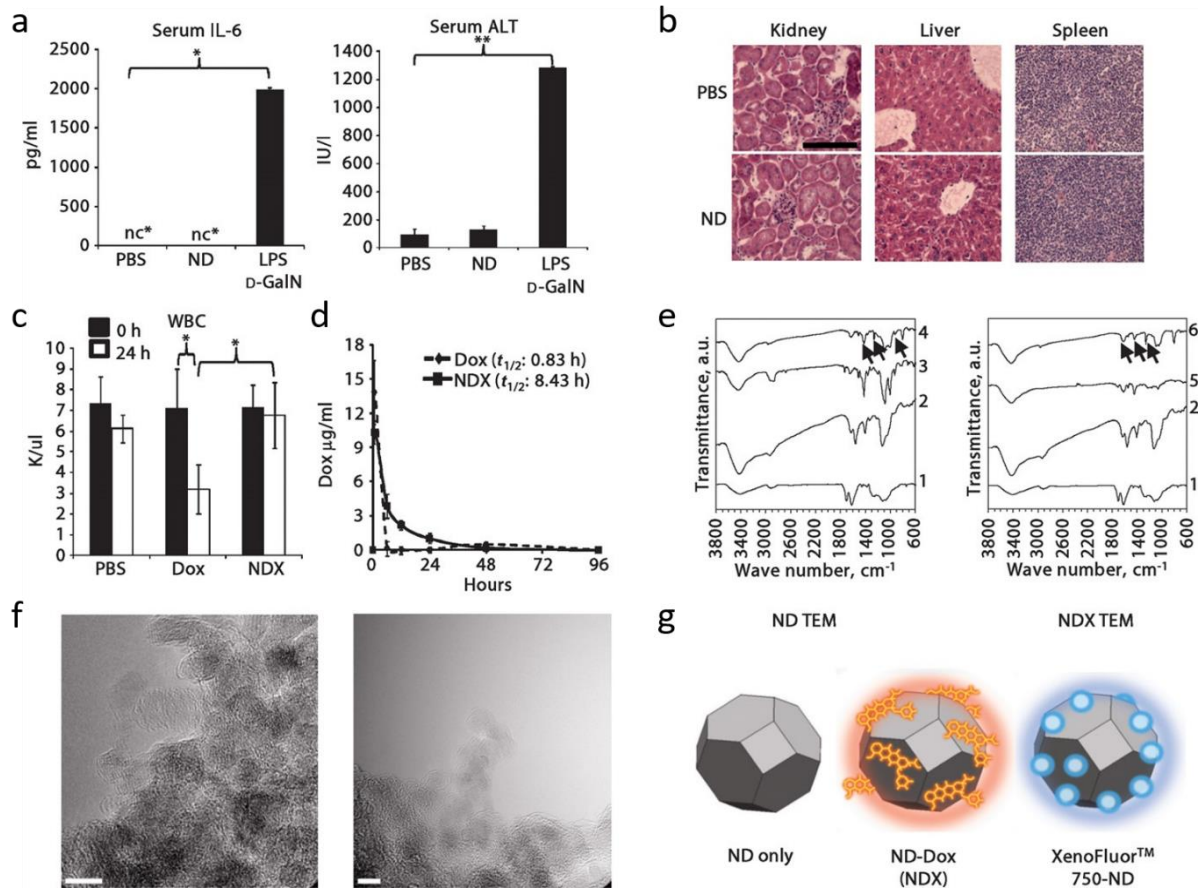

**Figure 2.** Nanodiamonds (NDs) are non-toxic and capable of conjugation with a variety of molecules. (a) Serum analysis of FVB/N mice tail vein injected with 500 μg of NDs (n = 3) or PBS (n = 3) for 1 week or lipopolysaccharide (LPS) (2.5 μg/kg)/D-galactosamine (D-GalN) (200 mg/kg) (n = 2) for 6 hours. nc, no change. Data are represented as means ± SD. *P < 0.006; **P < 0.001. (b) 40× hematoxylin and eosin (H&E) histopathological analysis of kidney, liver, and spleen tissue from treated mice. Scale bar, 100 μm. (c) Mean white blood cell (WBC) counts after treatment with PBS (n = 5), doxorubicin (Dox) (400 μg) (n = 5), or ND-conjugated Dox (NDX) (400 μg of Dox equivalent) (n = 5). Data are represented as means ± SD. *P < 0.002. (d) Blood circulation halftime (t1/2) analysis after treatment with Dox (200 μg) (n = 4)

or NDX (200 μg of Dox equivalent) (n = 4). Data are represented as means ± SD. (e) FTIR analysis of ND (spectra 1), ND-NH2 (spectra 2), free XenoFluor 750 dye (spectra 3), and XenoFluor 750–ND (spectra 4). Arrows denote the C-N stretch and N-H bend combination at 1261 cm$^{-1}$, benzene ring stretch at 1508 cm$^{-1}$, and vibration of aromatic C-H at 926 cm$^{-1}$. FTIR analysis of reduced ND (spectra 1), ND-NH2 (spectra 2), free Alexa Fluor 488 dye (spectra 5), and Alexa Fluor 488–ND (spectra 6). Arrows show IR features at 1261 cm$^{-1}$, which represent the C-N stretch and N-H bend in –CO–NH–C– groups, suggesting amide formation. Arrows also denote benzene ring stretch at 1619 and 1442 cm$^{-1}$. (f) Transmission electron microscopy (TEM) images of NDs and NDX. Scale bars, 5 nm. (g) Model of ND, NDX, and XenoFluor 750–ND. Adopted with permission from Chow *et al* [55].

## Diamond for Orthopaedics

Diamond is an emerging material in all areas of orthopaedics. Several factors should be considered when examining an orthopaedic implant material, such as surface topography and chemistry, roughness, wettability, and *in vitro* and *in vivo* cell growth. Due to the wearing issue, metal ion release and lack of cytocompatibility of the current metallic implants, diamond has been explored as an implant coating to overcome these issues. Another form of carbon structure is related to the composite-like characteristics of NCD grains embedded within an amorphous carbon matrix, known as 'nanostructured diamond' (NSD) or 'ultra-smooth nanostructured diamond' (USND) [23]. Catledge *et al.*,[56] and Chowdhury *et al.*,[57] investigated the application of NSD and UNSD for coating orthopaedic implant using CVD. The NSD diamond coatings are characterised as a mixture of sp$^3$- and sp$^2$-bonded carbon and nanometre-range grain size embedded in an amorphous carbon medium. The coatings typical encompass high hardness, excellent fracture toughness and low surface roughness properties on metallic surfaces. However, the coatings may delaminate [58-60] due to the thermal mismatch between the metals and diamond. By nature of the strong interfacial carbide layer that inherently forms during the deposition of diamond onto titanium surfaces, the tribological properties and coating adhesion are considered to be excellent [23, 56]. Cobalt chromium is commonly known for its use as high-wear bearings in the industry of orthopaedic implants. The high cobalt content of the alloy provides a stimulus for graphite to form as an interfacial layer. Similarly, transition metals such as iron or nickel play the same role due to the high solubility of carbon. The lack of an interlayer can result in poor coating quality and adhesion to the material.

To assess the capabilities of a material to facilitate osseointegration, Kokubo *et al.,*[61] developed a salt-based solution known as simulated body fluid (SBF) without the requirements of *in vitro* or *in vivo*. Fox *et al.,*[62] employed the SBF technique to demonstrate that apatite films could be deposited on PCD-coated silicon substrates after a 14-day incubation period. However, limitations are reported with diamond coatings on titanium substrates, for example the hydrogen embrittlement caused by residual stress in the coating process [58]. Although there is a large amount of residual stress in CVD diamond coatings, the coating adhesion may still be sufficient in the majority of the coated films due to bonded interfacial titanium carbide.

## Diamond surface characteristics (*in vitro*)

Surface topography and roughness

The surface topography and morphology of diamond vary according to its various fabrication methods. Mammalian cells respond differently to ordered or random patterns on a surface than to conventional, smooth or non-patterned surface topographies. The cell adhesion, proliferation and function on specifically designed surfaces can change the initial interaction, as well as the later developmental stages leading to tissue formation. As suggested by Webster *et al.,*[63] Catledge *et al.*,[23] and Grausova *et al.*,[64] cell adhesion is mainly dependent on the binding protein molecules excreted according to the surface roughness and morphology. Several features on a diamond surface may facilitate cell attachment. In this regard, a number of researchers are continuing to investigate the influences of surface topography on cell behaviour.

Surface roughness is typically measured; as a result, height profile on the surface of the material. A number of groups claim that increased surface roughness enhances the number of cells adhering to a substrate [65-67]. Nanoscale ceramics demonstrate high osteoblasts cell adherence, as reported on conventional alumina and titania [63]. Similarly, human osteoblast (MG63) cells cultured on titania (Ra = 40 nm) show greater cell viability and proliferation than on substrates with higher roughness (Ra = 100–170 nm) [68]. The more significant amount of viable cells adhering on substrates with nanoscale roughness may be due to a large number of atoms present on the surface, surface defects and grain boundaries [63]. These factors may influence the release of adhesion proteins from the cells to the substrate.

Currently, ND has not been investigated comprehensively enough to elucidate the roles of surface topography and roughness on individual particles. Nanodiamonds are thought to improve the mechanical properties of the scaffold, as illustrated by Fox *et al.,*[69] However*,* difficulties may arise with ND due to the agglomeration of particles. Zhang *et al.,*[47] have shown adding greater numbers of ND particles enhance biocompatibility with polymers, including poly-lactic acid (PLLA) (Fig. 3).

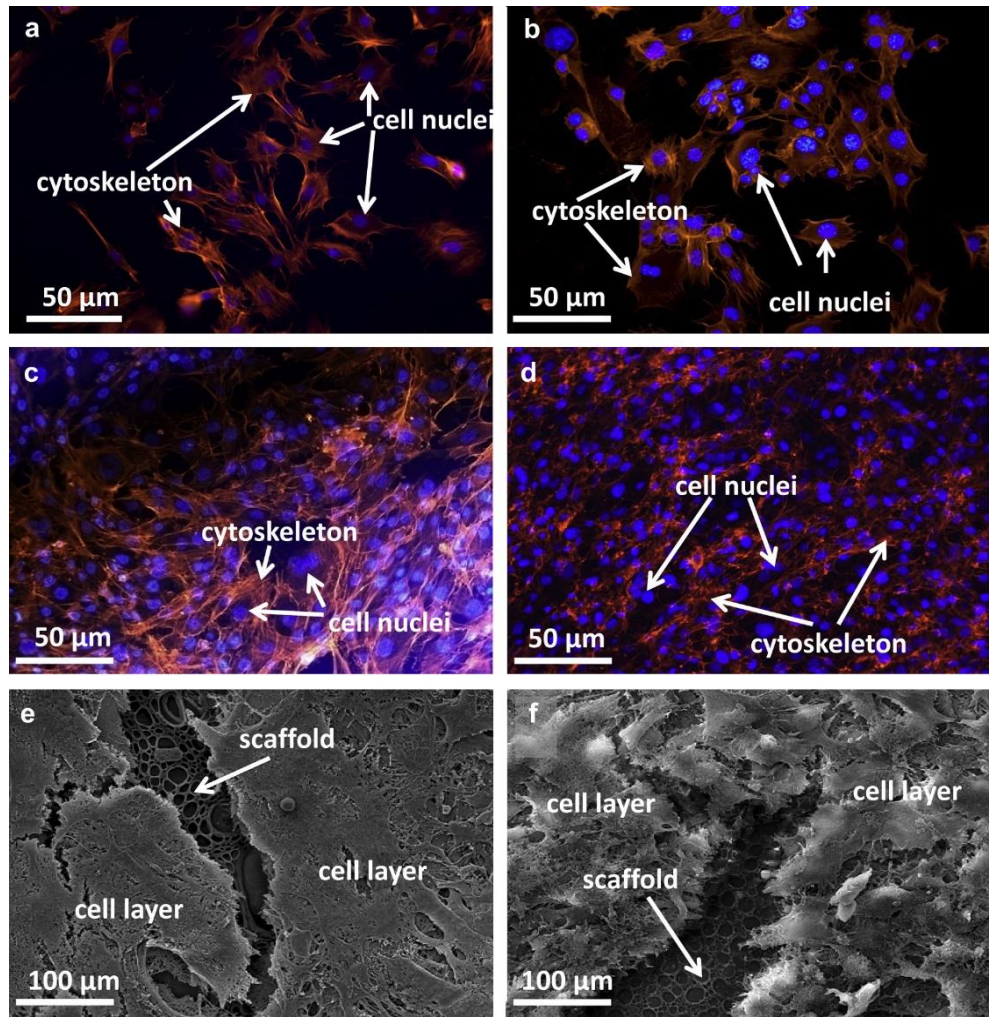

**Figure 3.** Morphology of 7F2 osteoblasts at 2 and 6 days of post-seeding on pure PLLA (a, c) and 10%wt ND-ODA/PLLA (b, d). Cell morphology of 2 days post-seeding on PLLA (a) and 10%wt ND-ODA/PLLA (b) shows that the attachment of osteoblasts on the surface of 10%wt ND-ODA/PLLA scaffold is as good as on the surface of a pure PLLA scaffold. Staining for nuclei is bis-benzimide (blue) and for actin cytoskeleton-phalloidin (red). SEM images of 7F2 grown on the different scaffolds after 6 days post-seeding: osteoblasts spread to confluence similarly on the surface of pure PLLA (e) and 10%wt ND-ODA/PLLA (f) scaffolds. Adopted with permission from Zhang *et al.,*[47].

Kalbacova *et al.,*[70] report that the highest initial SaOS-2 cell attachment, differentiation and biomineralization rates are observed on the 20 nm RMS scaffold amongst the surfaces that are fabricated with the same grain size and non-uniform roughness values of RMS = 20 nm, 270 nm and 500 nm. Similarly, Clem *et al.,*[5] showed that osteoblast adhesion and proliferation increase by application of USND. The ideal surface grain size and roughness remain controversial issues pursued by several research groups. However, some concluding remarks can be made in terms of the criteria of using surface topography and roughness for bone fixative implants. As reported by the vast majority of research on using diamond for bone growth, roughness levels of ≤ 100 nm and a grain size of approximately 2 μm is typically preferred for enhancing OB functions. A single value of diamond grain size and surface roughness cannot be justified for the cell response, as there is a counterbalance between surface topography and roughness.

Surface chemistry and wettability

Chemical vapour-deposited diamond is commonly hydrogen-terminated as-fabricated, leaving the surface to be partly hydrophobic. Clem *et al.,*[5] Lechleitner *et al*.,[51] and Yang *et al*.,[71] state that the water contact angle of these surfaces can range from 85 to 95°. To alter the wettability, the as-fabricated diamond film can be oxygen terminated, inducing hydrophilic properties at the surface. Studies by Tong *et al.,*[72] Kalbacova *et al.,*[70] and Lechleitner *et al*.,[51] report that oxygen plasma treatment increases the initial cell adhesion due to the enhancement of hydrophilic properties of the diamond surface. However, the literature reports mixed responses depending on the physical characteristics of the cells grown on the diamond. Lechleitner *et al*.,[51] confirm that hydrogen termination on NCD (hydrophobic) prevents renal epithelial cell (HK-2 cell line) adherence but oxygen-terminated NCD (hydrophilic) enhances HK-2 cell adherence and proliferation compared to that of borosilicate glass. The lack of certain functional polar groups, such as–OH and –COOH, may specifically inhibit the binding and subsequent growth of adherent cells. A hydrophilic diamond surface made *via* oxygen termination provides sites for the polar and ionic elements of the cell culture media to be situated, leading to the adherence of cells. In a study by Kloss *et al.,*[73] oxygen terminated NCD was implanted in sheep calvaria. The results showed increased levels of BMP-2 and higher numbers of OBs in closely bound contact with the implant in comparison to uncoated and hydrogen terminated scaffolds. The OB cells and extra-cellular matrix distinctively resembled those associated with early bone formation.

<u>Wear debris and bacterial infections</u>

The frictional force between the bone and implant can cause debris, and creates an inflammation which then results in pain and leads to aseptic loosening and implant failure [9, 11]. Such an occurrence can initiate the response of macrophages, which is dependent on the amount of wear debris and the size and concentration of the aggregated particles. Usually, large debris pieces are covered by fibrous tissue, whereas the smaller pieces are engulfed by macrophages and multinucleated giant cells. Phagocytic cells consume the wear debris and become activated release inflammatory cytokines and other factors that promote osteoclasts at the bone-implant interface [44, 74, 75].

A common modality for bacterial contamination or attachment of free-floating bacteria on the surface of the implant is during implant surgery. The bacteria, predominantly known for infections on orthopaedic devices, is *Staphylococcus aureus*. They can be dormant and then become active inside the body after implantation under the influence of body fluids. Once they become activated, these bacteria can colonise the implant surface and form biofilms. The biofilms are part of the inflammatory phase of the host's immune system which can result in implant failure [11, 76-78]. It is difficult to eradicate biofilms even though physical removal techniques and antibiotics are employed. To minimise these effects, diamond coatings have been explored as inhibitory coatings [21, 79-81].

A study by Wehling *et al.*,[46] highlights the importance of surface functionalisation of ND particles. It shows that oxygen containing functional groups and negatively charged surfaces have antibacterial properties (Fig. 4). Furthermore, they report that specific proteins control the effectiveness of the functional group by either covering or exposing them. In another study, Norouzi *et al.,*[82] discuss that the NDs inhibit bacterial adhesion due to the aggregation of the particles rather than an antibacterial property. They compare different types of NDs, showing the influence of particle size distribution. The surface morphology of the nanoparticles is shown to play a critical role in this case, inhibiting the activity of the bacteria around the aggregated NDs.

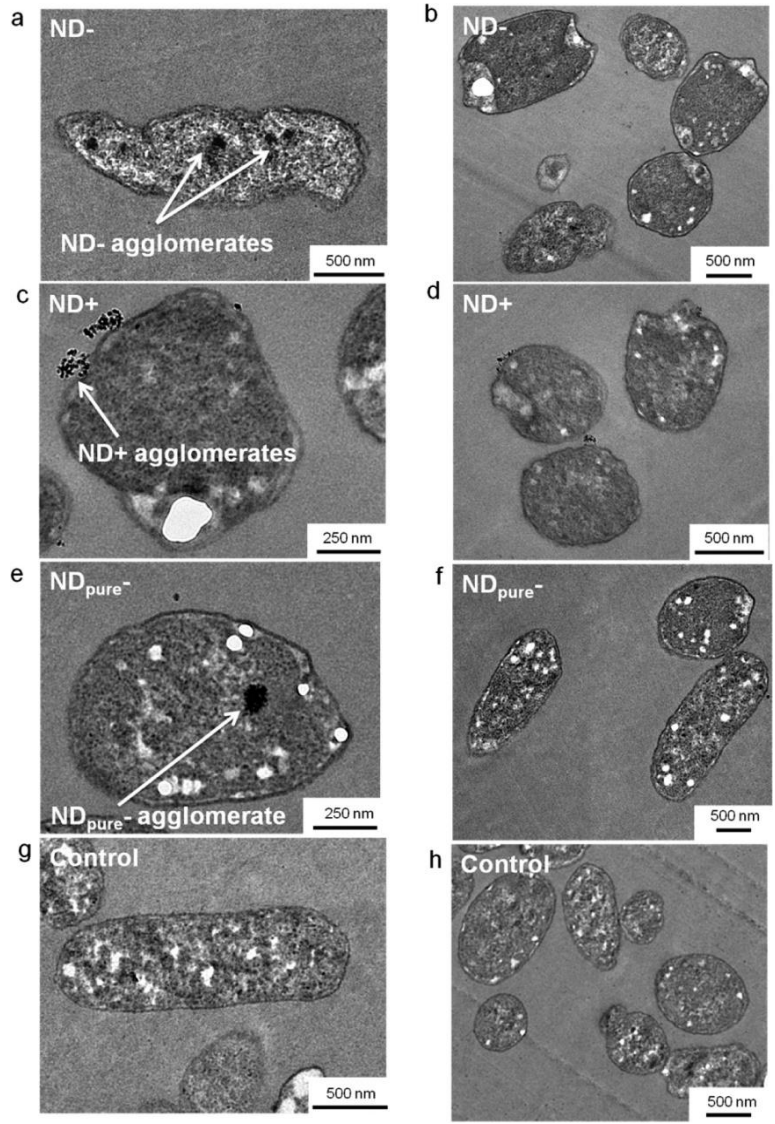

**Figure 4.** Transmission electron microscopy images of various types of nanodiamonds (NDs) in bacteria. The images indicate that, at sublethal ND concentrations of 0.5 mg/L, ND– is incorporated into E. coli cells and seems to deform the cellular shape (a, b). ND+ seems mainly to bind to cellular surface structures (c, d). Similar to ND–, agglomerates of negatively charged ND pure– are also found inside the cells, but they do not alter bacterial morphology (e, f), showing similar cell shapes to the ND-free control of E. coli (g, h). Adopted with permission from Wehling *et al*. [46].

## Biocompatibility of diamond (*in vivo*)

Despite the understanding of diamond under *in vitro* conditions, a thorough *in vivo* evaluation of diamond is still lacking. A work by Kloss *et al.,*[73] shows that the coating of endosseous implants with NCD allows stable functionalisation with oxygen termination (Fig. 5). The coating of the implant enhanced *in vivo* osseointegration in sheep calvaria. Likewise, Lemons *et al.,*[83] demonstrated the biocompatibility profile for bone integration with NSD on the distal femoral and proximal tibial bone in rabbits. The results showed better bone growth on the NSD coating than on uncoated Ti-6Al-4V. All the surfaces with exhibited trabecular, and compact bone are associated with the titanium alloy and diamond without any inflammation. Macro- and microscopically of both the titanium and diamond surfaces were histologically similar.

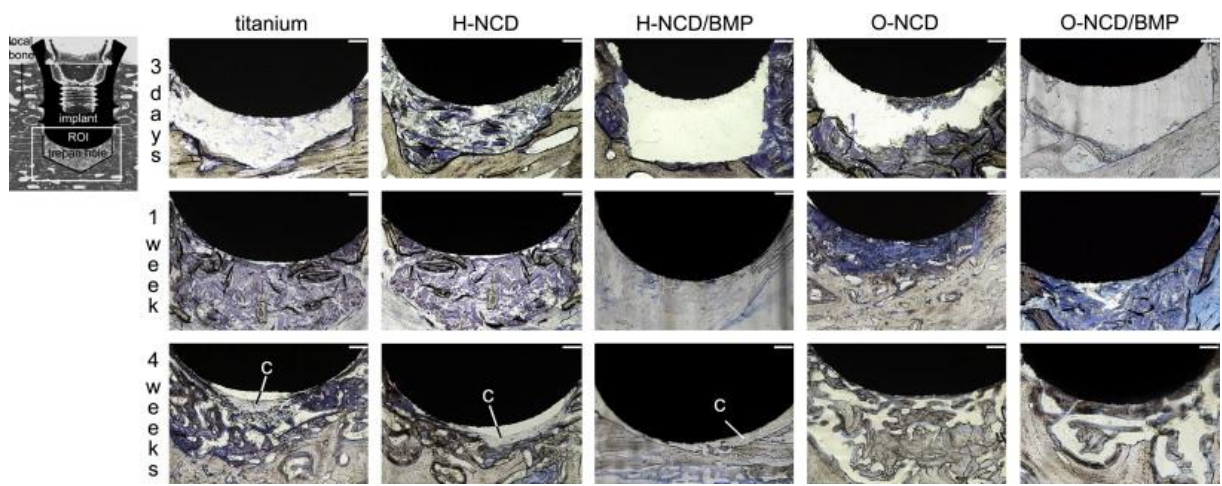

**Figure 5.** Histology of bone healing after implantation of BMP-2 coated NCD. According to the depicted scheme at the left side of the panel, representative examples of Toluidine Blue O stained bone sections are presented. Specimens from all experimental groups (3 days, 1 week and 4 weeks post-operation) were processed by the cutting–grinding method. Connective tissue (c) adjacent to the implant was observed at titanium, H–NCD and H–NCD/BMP. Bars indicate 200 μm. Adopted with permission from Kloss *et al.*[73].

More recently, Rifai *et al*., [79] established the capabilities of PCD coated additively manufactured titanium scaffolds in apatite-like conditions. The scaffolds were examined in a simulated body fluid, which was a protocol established by Kokubo *et al.,* [61] to analyse the bone-forming capabilities of a material *in vitro*. The results show that PCD-coated surfaces have a greater thickness apatite-like mineral layer compared to as-fabricated titanium or polished surfaces. Therefore, the PCD surfaces were found to be more suitable for the formation of bone than uncoated 3D-printed Ti-6Al-4V surfaces.

**Three-dimensional printing and diamond**

Additive manufacturing is renowned for introducing bespoke parts using computer-aided design. One of the more common uses of 3D-printing is with metals. The state-of-the-art of metal printing is with selective laser melting (SLM), which bonds the metal powder to complex structural forms (Fig. 6). The technique employs high-power density to print layer-by-layer into a complete net-shape of 99.9% relative density [84]. A variety of metals (such as stainless steel, cobalt chromium and titanium) can be used with SLM; however, they are limited by their ability to integrate with the bone and surrounding tissue [85]. The prospect of developing custom or patient-specific implants is attractive. Researchers are exploring this avenue to fabricate novel materials using 3D-printing, including diamond.

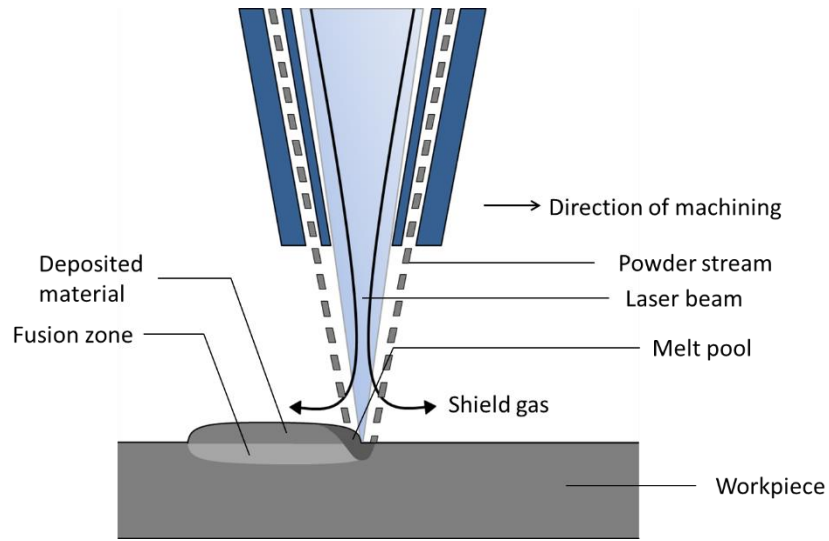

**Figure 6.** Schematic representation of metal 3D-printing using selective laser melting.

To achieve a 3D-printed diamond, understanding the feasibility and compatibility of 3D-printed metals is important. With the help of diamond particles, Rifai *et al*., [86] show that it is possible to coat NDs onto SLM titanium parts using a simple dip-coating method (Fig. 7). The study shows that 88% coverage can be achieved with improvements to the mammalian cell growth whilst synergistically reducing the bacterial adhesion. This phenomenon is particularly attributed to the wettability of the NDs, which is induced by the oxygen containing functional groups.

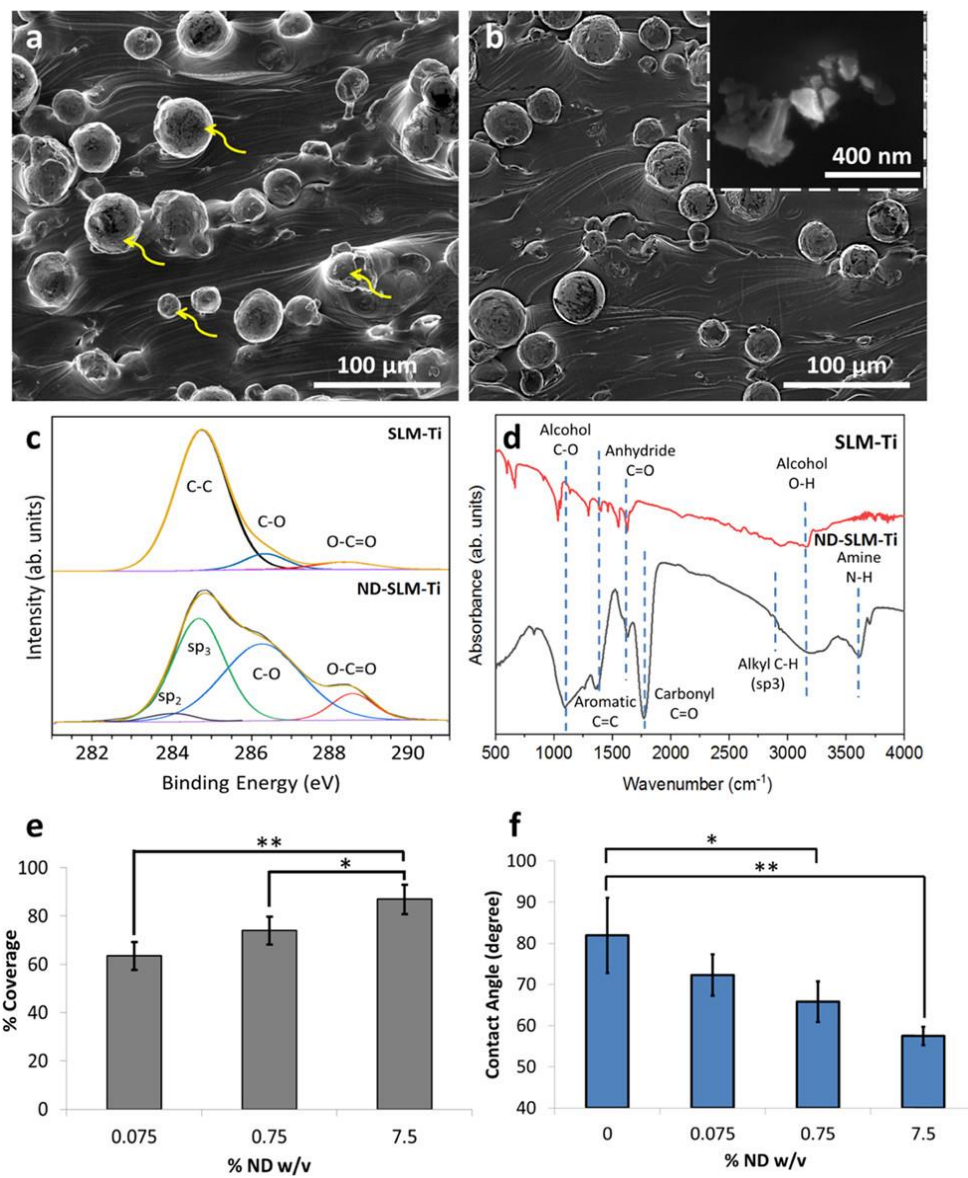

**Figure 7.** Surface characterization of nanodiamond (ND)-coated selective laser melted titanium (SLM-Ti) substrata (ND-SLM-Ti). (a) SEM image of the uncoated SLM-Ti substratum showing a large volume of partially melted particles (indicated by the yellow arrows). (b) SEM image of the ND-SLM-Ti substratum showing an inset of the ND attached to the surface of the SLM-Ti. (c) C 1s (carbon) high-resolution XPS spectra of uncoated SLM-Ti (top) showing peaks fitted for C–C, C–O, and O–C=O bonds. In contrast, as-received ND (bottom) shows peaks fitted for $sp^2$, $sp^3$, C–O, and O–C=O bonds. (d) FTIR spectra of uncoated SLM-Ti and as-received ND obtained using a potassium bromide pellet, detailing surface functional groups. Relative to SLM-Ti, ND adsorption regions indicate bonds of C–O, C=C,

C=O, $sp^3$, O–H, and N–H. (e) Surface coverage of ND over a $1 \times 1$ cm$^2$ dip-coated substratum in relation to the concentration of ND suspensions. Data = mean ± standard deviation. Coverage: * $p < 0.05$ and ** $p < 0.005$. (f) The water contact angle of the ND-coated SLM-Ti substratum. Data = mean ± standard deviation. Contact angle: * $p < 0.01$ and ** $p \leq 0.002$. Adopted with permission from Rifai *et al.*,[86].

At present, there are limitations with diamond being printed in one unit. Some researchers, including Rahmani *et al.,*[87] show that diamond can be fabricated using sintered diamond using SLM. These 3D-printed parts show improved wear resistance for potential mining applications. In another example, Roy *et al.,*[88] show that a titanium matrix can be embedded with diamond using a LENS™ system. The resultant structures are mechanically stable with a high Young's modulus, whereby 169±14 GPa and 629±102 GPa was achieved with 5% and 15% diamond respectively. More recently, Fox *et al.,*[89] show that diamond can be printed using direct energy deposition in a titanium composite. The protocol shows potential to build large (cm scale) structures with high mechanical and physical integrity. These substrates are also biocompatible with increasing concentrations of microdiamond showing enhance mammalian growth (Fig. 8). As such, 3D-printed diamond materials with excellent biocompatibility and mechanical integrity have a promising future in medicine.

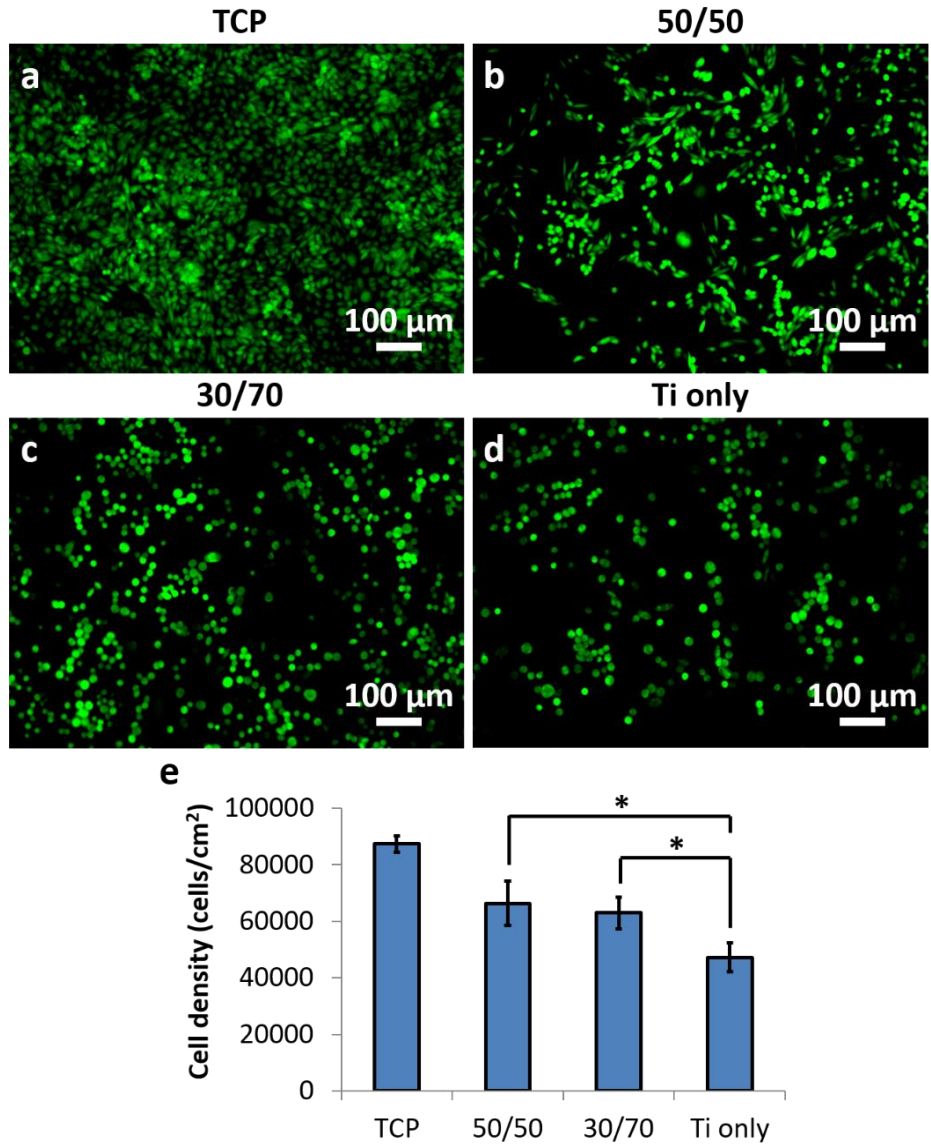

**Figure 8**. Fluorescent micrographs of Chinese Hamster Ovarian (CHO) cells grown over a period of 3 days. (a) tissue culture plastic TCP control, (b) 50/50 composite, (c) 30/70 composite, (d) as-fabricated titanium (Ti) samples (e) column graph showing the MTS cell density for the diamond–titanium (D–Ti) composites in comparison to the TCP and Ti control samples, averaged across three samples. In this case, the 50/50 and 30/70 samples show similar cell attachment between the two concentrations, though both D–Ti composites exhibit superior cell attachment compared to Ti. Data = mean ± standard deviation. ∗ Indicates statistical significance of $p \leq 0.05$. Adopted with permission from Fox *et al.,*[89].

In other examples, NDs can be printed using FDA approved filaments with a soft outer polymer shell. Studies by Fox *et al.,*[69] show the feasibility of NDs embedded in a poly-caprolactone (PCL) matrix. This material has the capabilities to enable tracking under the skin especially due to the fluorescent properties of the NDs. In a similar fashion, Houshyar *et al*.,[90, 91] also expand and explore the possibilities of NDs to be incorporated in PCL and polypropylene meshes. These scaffolds are versatile, where the NDs are employed to enhance mammalian cell growth, but also mechanically strengthen the polymer compared to traditional polymers. These polymer-based scaffolds typically find applications as a wound dressing or surgical meshes.

To produce a uniform diamond film on medical grade implant surfaces, Rifai *et al*.,[79] show that PCD can be coated on selective laser melted Ti-6Al-4V (Fig. 9a). Over the past couple of years, some diamond-based materials have come into existence in the market, including Carbodeon's microdiamond 3D-printed filament (Fig. 9b) [92]. This simple extrusion-based technique uses functionalised diamonds to increase the level of lubrication whilst printing at a speed of 500mm/s. This means that the nozzle will not wear over time. The overall mechanical properties of the material are enhanced with excellent stiffness, strength and adhesion between the printed layers. Likewise, Sandvik has developed one of first diamond and polymer composites using a slurry of diamond mixed in an epoxy resin and cured with ultraviolet light [93]. This composite has potential in a broad range of applications, with bespoke shapes and morphologies (Figs. 9c, d). Although these novel methods of 3D-printing diamond-based composites exist in the industry at this current time, they are not suitable for medical applications

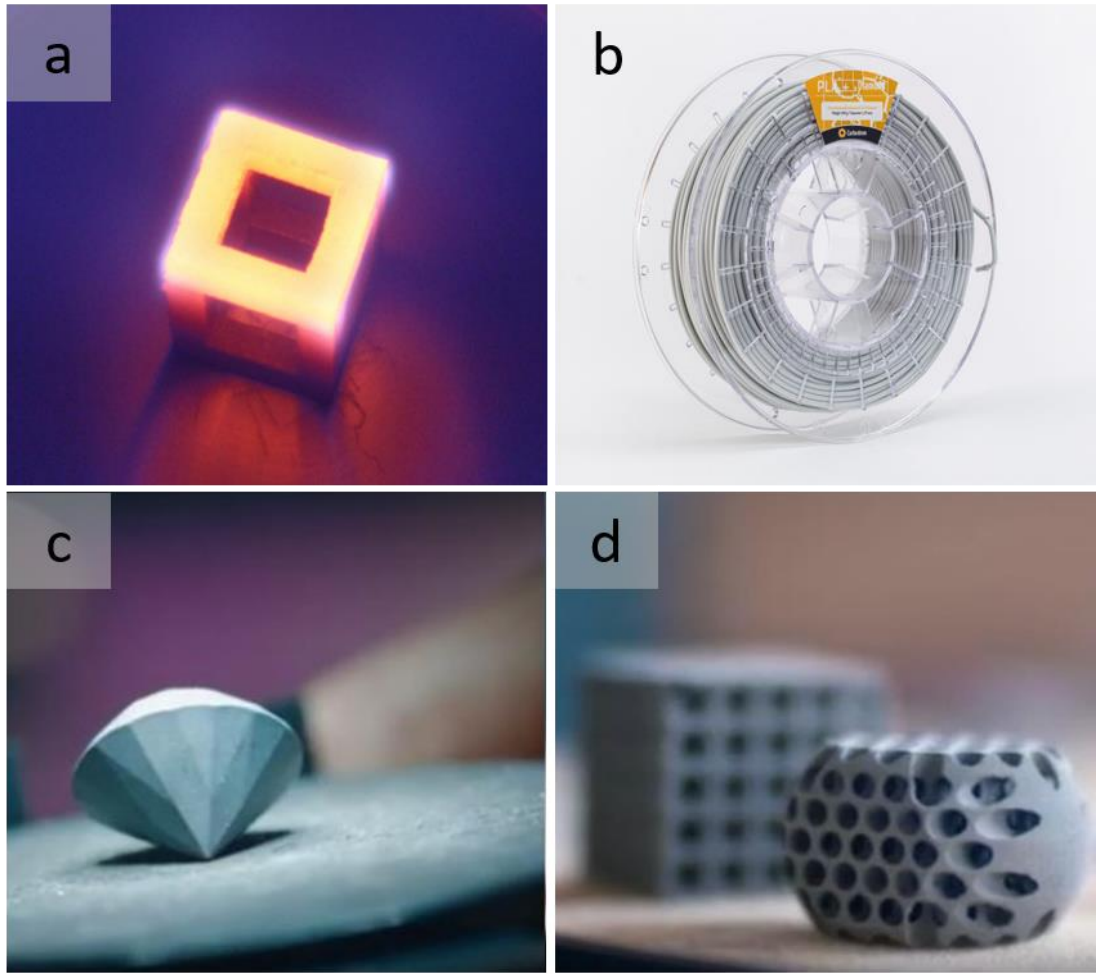

**Figure 9.** Forms of additively manufactured (3D-printed) diamond-based materials. (a) Selective laser melted titanium coated with chemical vapour deposition diamond. (b) Poly-lactic acid filament with embedded micro-diamond, manufacture by Carbodeon [92]. (c) Polymer and diamond composite fabricated using stereolithography by Sandvik [94]. (d) Another variation of Sandvik's unique production of diamond composites, showing the capabilities of creating porous materials [94].

# Future directions

In addition to the diamond-based implant fabrication techniques and biological assessments summarised in this review, the future trend of additively manufactured metal and polymers scaffolds are discussed. The potential capabilities of facile diamond coating on additively manufactured scaffolds can be improved further by optimising the protocols towards *in vivo* models and subsequent clinical trials. The diamond-coated substrates exhibited uniform coatings of PCD and ND, excellent cell viability and greater resistance to bacterial attachment than uncoated SLM-Ti substrates. This synergistic effect of increasing mammalian cell growth and reducing bacterial viability creates a number of new possibilities [79, 86]. Likewise, controlling the additive manufacturing process such as the line speed, rate, particle size, distribution, flowability and others can significantly impact the quality of the final structure. It is important that parametric optimisation is undertaken to construct an ideal material. In other instances, companies, including Sandvik, are leading the way from an industrial perspective by creating products with diamonds. They use stereolithographic methods to create customised epoxies embedded with diamonds [93]. Sandvik's process can be extracted from the polymer after printing and recycled, meaning that this approach is sustainable for the environment.

Although there are a number of potential avenues with 3D-printing of metals and polymers, the end-goal is to restore the targeted material in the body. When considering bone, for example, the mechanical properties and osteogenic properties of bone need to be matched for an ideal biomimetic cellular environment. As such, one of the long-standing problems in the modern era is to match the elastic modulus of bone with the fabricated material. Therefore, bioprinting soft materials could be the path that researchers take to provoke natural bone healing and remodelling with stimulating factors such as customised growth factors, nanoparticles or nanosystems.

**Conclusion**

Diamond is becoming extensively popular in the biomedical field, especially for orthopaedics. This review discussed diamond fabrication processes, surface characteristics, *in vitro* and *in vivo* assessments, and recent advancements in 3D-printing of diamond-based scaffolds. Its excellent biocompatibility, chemical stability and functionality make these scaffolds challenge the current state-of-the-art in biomaterial design and application. Modern synthesis techniques have increased research interest in diamond materials greatly. Although advances in using diamond for orthopaedic applications are increasing, gaps exist in terms of smart devices and patient-specific implants. While diamond can be a facilitator of cell growth or an enhancement to the mechanical integrity in hard-tissue applications, its ability to be incorporated in soft-materials could be an intriguing prospect for the future.

**Acknowledgements**

K.E.F. is supported by a CASS Foundation (Medicine and Science) Grant. K.E.F acknowledges the Clive and Vera Ramaciotti Foundation. K.E.F. acknowledges the Australian Research Council (ARC) for funding (IC160100026). S.H. acknowledges funding through the RMIT Vice-Chancellor's Research Fellowships.

**Figure Captions**

**Figure 1.** Schematic representation of two types CVD reactors. (a) hot filament and (b) plasma-enhanced microwave plasma reactor.

**Figure 2.** Nanodiamonds (NDs) are non-toxic and capable of conjugation with a variety of molecules. (a) Serum analysis of FVB/N mice tail vein injected with 500 μg of NDs (n = 3) or PBS (n = 3) for 1 week or lipopolysaccharide (LPS) (2.5 μg/kg)/D-galactosamine (D-GalN) (200 mg/kg) (n = 2) for 6 hours. nc, no change. Data are represented as means ± SD. *$P <$ 0.006; **$P < 0.001$. (b) 40× hematoxylin and eosin (H&E) histopathological analysis of kidney, liver, and spleen tissue from treated mice. Scale bar, 100 μm. (c) Mean white blood cell (WBC) counts after treatment with PBS (n = 5), doxorubicin (Dox) (400 μg) (n = 5), or ND-conjugated Dox (NDX) (400 μg of Dox equivalent) (n = 5). Data are represented as means ± SD. *$P <$ 0.002. (d) Blood circulation halftime (t1/2) analysis after treatment with Dox (200 μg) (n = 4) or NDX (200 μg of Dox equivalent) (n = 4). Data are represented as means ± SD. (e) FTIR analysis of ND (spectra 1), ND-NH2 (spectra 2), free XenoFluor 750 dye (spectra 3), and XenoFluor 750–ND (spectra 4). Arrows denote the C-N stretch and N-H bend combination at 1261 cm−1, benzene ring stretch at 1508 cm−1, and vibration of aromatic C-H at 926 cm−1. FTIR analysis of reduced ND (spectra 1), ND-NH2 (spectra 2), free Alexa Fluor 488 dye (spectra 5), and Alexa Fluor 488–ND (spectra 6). Arrows show IR features at 1261 cm−1, which represent the C-N stretch and N-H bend in –CO–NH–C– groups, suggesting amide formation. Arrows also denote benzene ring stretch at 1619 and 1442 cm−1. (f) Transmission electron microscopy (TEM) images of NDs and NDX. Scale bars, 5 nm. (g) Model of ND, NDX, and XenoFluor 750–ND. Adopted with permission from Chow *et al*.[55]

**Figure 3.** Morphology of 7F2 osteoblasts at 2 and 6 days of post-seeding on pure PLLA (a, c) and 10%wt ND-ODA/PLLA (b, d). Cell morphology of 2 days post-seeding on PLLA (a) and 10%wt ND-ODA/PLLA (b) shows that the attachment of osteoblasts on the surface of 10%wt ND-ODA/PLLA scaffold is as good as on the surface of a pure PLLA scaffold. Staining for nuclei is bis-benzimide (blue) and for actin cytoskeleton-phalloidin (red). SEM images of 7F2 grown on the different scaffolds after 6 days post-seeding: osteoblasts spread to confluence similarly on the surface of pure PLLA (e) and 10%wt ND-ODA/PLLA (f) scaffolds. Adopted with permission from Zhang *et al.*[47]

**Figure 4.** Transmission electron microscopy images of various types of nanodiamonds (NDs) in bacteria. The images indicate that, at sublethal ND concentrations of 0.5 mg/L, ND– is incorporated into E. coli cells and seems to deform the cellular shape (a, b). ND+ seems mainly to bind to cellular surface structures (c, d). Similar to ND–, agglomerates of negatively charged ND pure– are also found inside the cells, but they do not alter bacterial morphology (e, f), showing similar cell shapes to the ND-free control of E. coli (g, h). Adopted with permission from Wehling *et al.*[46]

**Figure 5.** Histology of bone healing after implantation of BMP-2 coated NCD. According to the depicted scheme at the left side of the panel, representative examples of Toluidine Blue O stained bone sections are presented. Specimens from all experimental groups (3 days, 1 week and 4 weeks post-operation) were processed by the cutting–grinding method. Connective tissue (c) adjacent to the implant was observed at titanium, H–NCD and H–NCD/BMP. Bars indicate 200 μm. Adopted with permission from Kloss *et al.*[73]

**Figure 6.** Schematic representation of metal 3D-printing using selective laser melting.

**Figure 7.** Surface characterization of nanodiamond (ND)-coated selective laser melted titanium (SLM-Ti) substrata (ND-SLM-Ti). (a) SEM image of the uncoated SLM-Ti substratum showing a large volume of partially melted particles (indicated by the yellow arrows). (b) SEM image of the ND-SLM-Ti substratum showing an inset of the ND attached

to the surface of the SLM-Ti. (c) C 1s (carbon) high-resolution XPS spectra of uncoated SLM-Ti (top) showing peaks fitted for C–C, C–O, and O–C=O bonds. In contrast, as-received ND (bottom) shows peaks fitted for $sp^2$, $sp^3$, C–O, and O–C=O bonds. (d) FTIR spectra of uncoated SLM-Ti and as-received ND obtained using a potassium bromide pellet, detailing surface functional groups. Relative to SLM-Ti, ND adsorption regions indicate bonds of C–O, C=C, C=O, $sp^3$, O–H, and N–H. (e) Surface coverage of ND over a $1 \times 1$ cm$^2$ dip-coated substratum in relation to the concentration of ND suspensions. Data = mean ± standard deviation. Coverage: * $p < 0.05$ and ** $p < 0.005$. (f) The water contact angle of the ND-coated SLM-Ti substratum. Data = mean ± standard deviation. Contact angle: * $p < 0.01$ and ** $p \leq 0.002$. Adopted with permission from Rifai *et al.*[86]

**Figure 8**. Fluorescent micrographs of Chinese Hamster Ovarian (CHO) cells grown over a period of 3 days. (a) tissue culture plastic TCP control, (b) 50/50 composite, (c) 30/70 composite, (d) as-fabricated titanium (Ti) samples (e) column graph showing the MTS cell density for the diamond–titanium (D–Ti) composites in comparison to the TCP and Ti control samples, averaged across three samples. In this case, the 50/50 and 30/70 samples show similar cell attachment between the two concentrations, though both D–Ti composites exhibit superior cell attachment compared to Ti. Data = mean ± standard deviation. ∗ Indicates statistical significance of $p \leq 0.05$. Adopted with permission from Fox *et al.*[89]

**Figure 9.** Forms of additively manufactured (3D-printed) diamond-based materials. (a) Selective laser melted titanium coated with chemical vapour deposition diamond. (b) Poly-lactic acid filament with embedded micro-diamond, manufacture by Carbodeon.[92] (c) Polymer and diamond composite fabricated using stereolithography by Sandvik.[94] (d) Another variation of Sandvik's unique production of diamond composites, showing the capabilities of creating porous materials.[94]